\documentclass[english]{article}
\usepackage[T1]{fontenc}
\usepackage[latin9]{inputenc}
\usepackage{geometry}
\usepackage{array}
\usepackage{booktabs}
\usepackage{multirow}
\usepackage{amsmath}
\usepackage{graphicx}
\usepackage{setspace}

\makeatletter

\providecommand{\tabularnewline}{\\}

\makeatother

\usepackage{babel}
\begin{document}
\title{Entropic Spatial Auto-correlation of Voter Uncertainty and Voter Transitions
in Parliamentary Elections}
\author{Omar El Deeb\thanks{Department of Natural Sciences, Lebanese American University}}
\maketitle

\begin{abstract}
This paper studies a novel spatial auto-correlation model of voter
uncertainty across districts. We use the Moran $I$ index to measure
the auto-correlation of Shannon, relative Shannon, Tsallis and relative
Tsallis entropies of regional electoral outcomes with respect to geographic
adjacency, proximity and sectarian adjacency. Using data from the
Lebanese parliamentary elections, we find strong geographic and gravitational
adjacency correlations. More importantly, there is a notably strong
correlation in sectarian adjacency in both $2018$ and $2022$ elections,
with a very high level of confidence. This result asserts the dominance
of the sectarian factor in Lebanese politics. We also introduce the
method of maximized general entropy estimation that allows us to determine
the Markov transition matrix of voters between consecutive elections.
\end{abstract}

\section{Introduction}

Entropy is a fundamental concept in thermodynamics and statistical
physics that measures disorder or randomness in a system. In thermodynamics,
it is a measure of the amount of energy that is unavailable to do
work. In information theory, it is a measure of the amount of uncertainty
or randomness in a message. Entropy has applications in a wide range
of fields, including physics, chemistry, biology, and engineering.
In physics, entropy is used to explain the behavior of gases, as well
as the behavior of energy and matter on a molecular level. In chemistry,
entropy is used to predict the direction of chemical reactions and
the distribution of energy among particles in a system. In biology,
entropy is used to study the behavior of complex systems, such as
the evolution of species while engineering, entropy is used to design
efficient systems for storing, transmitting, and processing information.
It has been also widely employed in multidisciplinary fields ranging
from information sciences, machine learning, social sciences and economics
to astrophysical systems, heavy ion collisions, election studies and
voter transitions \cite{Econ1, EntropyBook, Antweiler, Applications, Socio-economic, Elections1, Elections2, HeavyIon}.

In election studies, entropy can be used to measure the degree of
uncertainty or randomness in an election. This could be useful for
predicting the outcome of an election or for analyzing the effectiveness
of different voting systems. Entropy can also be used to measure the
diversity of preferences among voters, which could be useful for identifying
trends or patterns in voter behavior. Additionally, entropy can be
used to investigate the fairness of an election, by comparing the
distribution of votes across different candidates, parties or coalitions.
Particularly in elections, higher entropy signifies a more uniform
spread of votes to various lists, while lower entropy indicates more
concentration of votes for less lists. A maximum entropy indicates
equal distribution of votes for all lists while a zero entropy implies
that one list has one all of the votes.

After it was initially proposed by Claude Shannon in \cite{Shannon},
the entropy equation was heavily and successfully used in information
theory and all subsequent related fields. Other formulations of entropy
measures were later proposed and implemented in social systems, like
the Tsallis entropy. Tsallis entropy and Shannon entropy are both
measures of entropy, but they are based on different mathematical
definitions and have different applications. In general, Tsallis entropy
is more versatile and applicable to a wider range of systems than
Shannon entropy \cite{Tsallis1, Tsallis2}. Shannon's entropy in this
regard could be considered as a special case of the more general Tsallis
entropy \cite{Generalization}. 

An important tool that we utilize in this paper is the spatial auto-correlation
measure known by Moran's $I$ index of a spatial distribution \cite{Moran}.
It is an inferential statistic used to measure the spatial auto-correlation
between locations and feature values simultaneously \cite{Moran1, Moran2}.
The Moran index is a measure of spatial auto-correlation, which is
the degree to which the values of a variable are clustered together
in space. It can be used to analyze a wide range of spatial data,
including data on population density, economic activity, and environmental
factors. It is calculated by comparing the observed values of a variable
to the values that would be expected if the variable were randomly
distributed in space. A positive Moran index indicates that similar
values tend to be located near each other, while a negative Moran
index indicates that dissimilar values tend to be located near each
other. The Moran index is commonly used in geography, sociology, and
other fields to study the patterns and trends in spatial data \cite{Moran3, Moran4}.

Entropy maximization is the process of finding the distribution of
a random variable that has the maximum entropy, subject to certain
constraints. By finding the distribution with the maximum entropy,
researchers can make predictions about the behavior of the system
and its properties. It is a powerful tool for analyzing complex systems
and predicting their behavior. Entropy maximization is applied in
various fields, such as information theory, where it can be used to
design efficient data compression algorithms or to study the reliability
of communication systems. Entropic models were heavily used in electoral
studies to analyze cases as the uncertainty among voters and correlations
between increased uncertainty and voter turnout. The method of entropy
maximization was also employed to analyze voter transitions between
consecutive elections \cite{EntMax2, EntMax3, EntMax4, EntMax5}. Voter
transition refers to the process of voters switching their support
from one candidate or political party to another. This can happen
for many reasons, such as changes in the political landscape, shifts
in voter attitudes and priorities, or the influence of external factors,
such as the media or campaign tactics. Voter transitions are a key
aspect of elections and political analysis, as they can help to predict
the outcome of an election and to understand the factors that drive
voter behavior. In general, voter transitions are studied using techniques
from political science and sociology, such as survey research, statistical
analysis, and network analysis. These methods are used to identify
trends and patterns in voter behavior and to predict the outcome of
an election based on voter preferences and voting patterns \cite{Transition1, Transition2, Transition3, Transition4}.

Spatial auto-correlation methods were used in numerous fields of research
including geography, plant populations, traffic crashes, landscape
patterns, drought propagation, satellite data, migration flows, tectonics,
image segmentation, COVID-19 spread patterns, thermal patterns and
in almost all fields of applied sciences \cite{Applications1,Applications2,Applications3,Applications4,Applications5,Applications6,Applications7,Applications8,Applications9,Applications10}.
Many models from statistical physics were used in the analysis of spatial relations between electoral turnouts and outcomes \cite{Borghesi, Mobility, Vmodel, Abstention}, however, there is a gap in the literature in utilizing spatial auto-correlation techniques in the study of entropy distribution in electoral outcomes. 

In this paper, we develop a novel entropic spatial auto-correlation
method by applying the Moran's index of spatial auto-correlation on
various entropic measures across districts. This method allows us
to be able to spatially relate voter behavior across districts or
across other feature values. We apply the method on the outcomes of
the Lebanese parliamentary elections and we use it as a demonstrational
protocase. The most recent two parliamentary elections were held in
May, 2018 and May, 2022 respectively, adopting same electoral law,
district division and seat allocation. The country is divided into
$15$ electoral areas comprising $26$ regional districts, and $128$
seats are allocated at the district level. Lebanon is a democratic
country and holds regular parliamentary elections. The political system
of Lebanon is based on a multi-party system, with several political
parties representing different ideologies and interest groups. The
parliament of Lebanon is made up of 128 members, who are elected through
a system of proportional representation. Overall, the democratic elections
in Lebanon are an important part of the country's political system
and its commitment to democratic principles. Several studies have
qualitatively investigated some recent elections in Lebanon \cite{LebElec1, LebElec2, LebElec3, LebElec4}
but there is a huge gap in quantitative analysis of its elections
in general and very scarce academic studies about the latest couple
of them. 

Entropic measures provide a way to quantify how evenly the votes are distributed among different lists in a district. A greater level of entropy indicates greater uniformity and distribution of votes among the lists. This means that a higher entropy measure can indicate a shift in electoral dominance from fewer lists to a more even distribution among multiple lists. This measure of entropy also indicates that there is more uncertainty in the results. The spatial correlation of entropy measures proves that electoral voting behaviors are strongly correlated between adjacent districts. Spatial correlation has been applied to other schemes than elections in the literature. However, our novel contribution is to specifically apply these techniques on entropic measures on the district level. This combination of two known techniques creates a new tool for the spatial study of voting behavior. We also introduce the concept of sectarian adjacency to inspect the effect of this factor on voting outcomes. Furthermore, we apply concept of gravitational adjacency which is commonly used in economics and econophysics models, to analyze correlations of voting outcomes.

Using the methods of entropy and spatial auto-correlations, we show
that the uncertainty outcomes across districts are spatially auto-correlated
by geographic adjacency, voter gravitational adjacency and most apparently
by sectarian adjacency. Voters from physically neighboring districts
mostly have correlated voting uncertainty patterns. Voters from regions
with higher numbers of voters and closer distances between their centers
mostly have correlated voting uncertainty patterns. We also show that
voters from districts that have similar sectarian majorities have
correlated voting uncertainty patterns.

In this paper, we also study voter transitions among lists between
$2018$ and $2022$ parliamentary election, and determine the Markov
transition matrices of voting outcomes in some selected districts.
We make use of the significant advancement in the theoretical understanding
of the generalized maximum entropy (GME) estimator in recent years.
Additionally, the use of the GME estimator has become more widespread
with the use of dedicated packages in statistical or computational
software like in SAS, MATLAB, STATA,.... \cite{SAS, STATA, GMEEstimator}.
We employ the GME estimator package in SAS in order to estimate voter
transitions among lists between the $2018$ and $2022$ Lebanese parliamentary
election.

The structure of the paper is as follows: In section 2, we introduce
Shannon's and Tsallis' entropy, Spatial auto-correlation and Moran's
Index and introduce our model. We also define our methods used in
voter migration analysis. In section 3, we present our results and
discuss them, then we conclude in section 4. 

\section{Model}

\subsection{Shannon's and Tsallis' Entropy}

Entropy is a measure of a state disorder. In elections outcomes, entropy
represents the dispersion of the voting patterns for different lists,
parties or coalitions. We use the definition of Shannon's entropy,
as defined in information theory to be

\begin{equation}
H_{j}=-\sum_{i=1}^{N_{j}}p_{i,j}\log(p_{i,j})
\end{equation}

where $H_{j}$ is Shannon's entropy in district $j$, $p_{i,j}$ is
the relative number of votes of list $i$ in district $j$ and $\sum_{i=1}^{N_{j}}p_{i,j}=1$
and $N_{j}$ is the number of lists in the $j$-th district. It is
a non-negative measure of uncertainty. Shannon's entropy is an extensive
quantity, as the entropies are additive. The maximum entropy value
occurs when the distribution is uniform or when all events are equiprobable,
hence when $p_{i,j}=\frac{1}{N_{j}}$ while its minimum occurs when
there is certainty about one particular event, corresponding to a
single list gaining $100\%$ of the votes in its district. In this
sense, 
\begin{equation}
H_{\max_{j}}=-N_{j}\log\left(\frac{1}{N_{j}}\right)=N_{j}\log N_{j}
\end{equation}

We define the relative entropy $H_{R_{j}}$ of the vote distribution
in a district $j$ as the normalized Shannon's entropy with respect
to its maximal value $H_{\max_{j}}$, hence it is given by

\begin{equation}
H_{R_{j}}=\frac{H_{j}}{H_{\max_{j}}}=-\frac{\sum_{i=1}^{N_{j}}p_{i,j}\log(p_{i,j})}{N_{j}\log N_{j}}
\end{equation}

\begin{figure}
\begin{centering}
\includegraphics[scale=0.26]{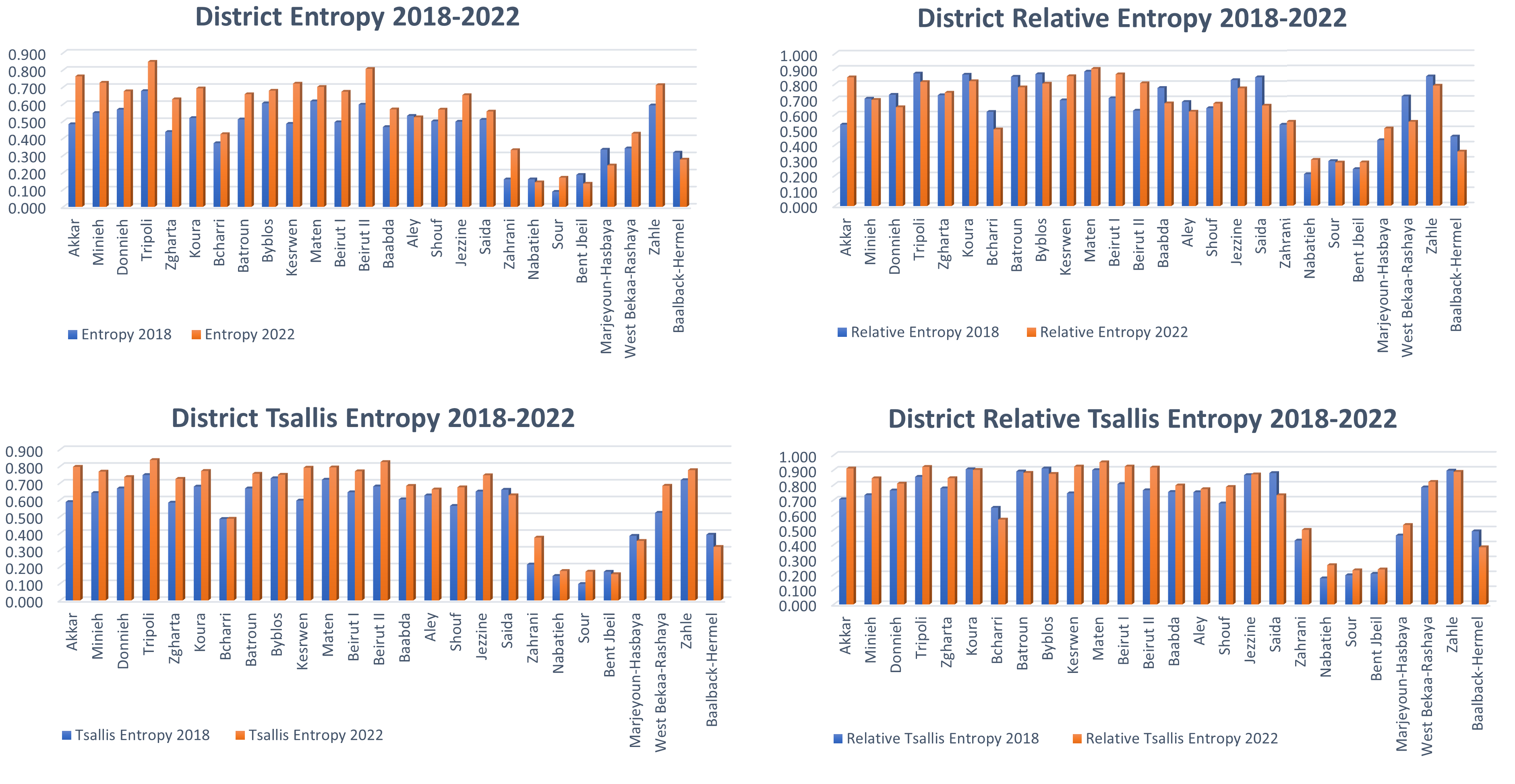}
\par\end{centering}
\centering{}\caption{A summary of the entropy data of the Lebanese Parliamentary elections
in 2018 and 2022. The upper figures show Shannon's Entropy and Relative
Shannon's entropy in each electoral district, while the lower figures
show the values of Tsallis and relative Tsallis' entropies of entropic
index $2$ in the same districts.}
\end{figure}

An alternative non-additive description of entropy is that defined
by the Tsallis entropy, and it is widely used in the study of complex
system. The Tsallis entropy of a system on $N_{j}$ events is given
by:

\begin{equation}
S_{q_{j}}=-\frac{1-\sum_{i=1}^{N_{j}}p_{i,j}^{q}}{1-q}
\end{equation}

where the $q$ is the entropic index, and can be any real parameter.
In contrast with Shannon's entropy, Tsallis entropy is a non-extensive
quantity as entropies of different systems are not additive according
to equation (4). In the limit $q\rightarrow1$, Tsallis entropy reduces
to the Shannon's entropy defined in (1). The maximum value of $S_{q_{j}}$
occurs when all events are equiprobable so that

\begin{equation}
S_{q\max_{j}}=-\frac{1-N_{j}^{1-q}}{1-q}
\end{equation}

and the relative Tsallis entropy is:

\begin{equation}
S_{qR_{j}}=\frac{1-\sum_{i=1}^{N_{j}}p_{i,j}^{q}}{1-N_{j}^{1-q}}
\end{equation}

Choosing the entropic index in Tsallis entropy involves considering various factors, such as the nature of the system under study, the characteristics of the data, and the desired properties of the entropy measure. For $q>1$, Tsallis entropy is more sensitive to rare events: Increasing the value of $q$ makes Tsallis entropy more sensitive to rare events in the system, however, there is no universal best value for $q$.
In many cases, the value of $q=2$ has been found to be useful in describing the statistical properties of systems. In complex systems, the value of $q=2$ was found to be particularly useful in capturing the hierarchical structure of networks and biological systems. More importantly, in nonlinear dynamical systems, it was found to be useful in characterizing the long-range correlations and spatiotemporal complexity \cite{Qvalue1, Qvalue2, Qvalue3, Qvalue4}. In this paper, we study Tsallis entropy with entropic index $q=2$, hence we specifically use 

\begin{equation}
S_{j}=1-\sum_{i=1}^{N_{j}}p_{i,j}^{2}
\end{equation}

and its normalized relative value

\begin{equation}
S_{R_{j}}=\frac{1-\sum_{i=1}^{N}p_{i,j}^{2}}{1-\frac{1}{N_{j}}}
\end{equation}

\subsection{Spatial auto-correlation}

In this paper, we introduce the idea of measuring the spatial auto-correlation
of election entropies across districts in relation to their geographic
adjacency, and the driving distance between their centers.

Moran's $I$ index is an inferential statistic used to measure the
spatial auto-correlation based both on locations and feature values
simultaneously. It is given by:

\begin{equation}
I=\frac{N\Sigma_{ij}W_{ij}(X_{i}-\bar{X})(X_{j}-\bar{X})}{\Sigma_{ij}W_{ij}\Sigma_{i}(X_{i}-\bar{X})^{2}}
\end{equation}

where $W_{ij}$ resembles a row-standardized location value matrix. 

We consider $3$ different parameterizations corresponding to geographic
adjacency, gravitational voter's weight and sectarian adjacency between
districts. They are defined as follows:
\begin{itemize}
\item First, $W_{ij}$ is taken to be the geographic adjacency matrix between
districts. Based on the district division in an election, districts
that share a common border are defined to be adjacent such that: 
\begin{equation}
W_{ij}=\begin{cases}
1 & \text{if districts \ensuremath{i\text{ and \ensuremath{j} are adjacent}}}\\
0 & \text{Otherwise}
\end{cases}
\end{equation}
\item Second, $W_{ij}$ is defined in terms of the gravitational voter weight
which is directly proportional to the numbers of voters $v_{i}$ and
$v_{j}$ and inversely proportional to the square
of the driving distances $d_{ij}$ between the centers of districts
$i$ and $j$. This leads to:
\end{itemize}
\begin{equation}
W_{ij}=\frac{v_{i} v_{j}}{d_{ij}^{2}}
\end{equation}

\begin{itemize}
\item Third, we set $W_{ij}$ to be an adjacency matrix that relates districts
according to the sectarian majority in each. In this regard, two districts
are considered adjacent if they have the same majority sect and non-adjacent
otherwise. 
\[
W_{ij}=\begin{cases}
1 & \text{if districts \ensuremath{i\text{ and \ensuremath{j} have same majoirty sect}}}\\
0 & \text{Otherwise}
\end{cases}
\]
\end{itemize}
In all these cases, the diagonal terms are set to be zero, $W_{ii}=0$,
and the rows are standardized such that $\sum_{j=1}^{N_{j}}W_{ij}=1\ \forall i$. 

In equation (9), $N$ is the number of regions under consideration
and $X_{i}$ and $X_{j}$ represent the entropies in regions $i$
and $j$ respectively. $\bar{X}$is the average entropy, and it is
given by $\bar{X}=\frac{\Sigma_{i}X_{i}}{N}$. We consider four different
parameterizations of $X$ corresponding to Shannon's entropy $H_{j}$,
relative Shannon's entropy $H_{R}$, Tsallis entropy $S$ (of entropic
index $2$) and its relative value $S_{R}$ defined in equations (1),
(3), (7) and (8) respectively.

The numerical outcome of $I$ falls between $-1$ and $+1$ and it
indicates whether a distribution is dispersed, random or clustered.
A value of $I$ close to $0$ indicates a random distribution, while
positive values indicate clustered spatial distribution and negative
values indicate dispersion. Larger values of $|I|$ nearer to 1 mean
stronger clustering (positive I) or stronger dispersion (negative
I). 

The $z_{I}$-score associated to these statistics is defined by:

\begin{equation}
z_{I}=\frac{I-E[I]}{\sqrt{V[I]}}
\end{equation}

where $E[I]$ is the expected value and $V[I]$ is the variance, which
are defined explicitly in the Appendix. 

The $z$-score or its corresponding $p$-value of the statistic are
used to reject the null hypothesis and eliminate the possibility of
a random pattern leading to the obtained value of the Moran $I$ statistic.
In this paper, we take a $95\%$ confidence level corresponding to
$z_{I}>1.96$ or equivalently to $p<0.05$ in order to confirm the
outcome of clustering or dispersion of our electoral spatial data
under consideration. When the $p$-value is statistically significant,
and based on the value of $I$ we can determine the pattern of the
distribution.

\subsection{Voter Transitions}

We use the method of "maximum entropy'' in order to estimate the
voter transitions among lists between consecutive elections. The maximum
entropy estimator utilizes a set of imposed constraints and available
information in order to find the most uniform estimate the maximizes
the entropy.
\begin{center}
\begin{table}
\centering{}%
\begin{tabular}{c>{\centering}b{2cm}>{\centering}m{2cm}>{\centering}m{2cm}>{\centering}m{2cm}>{\centering}m{2.5cm}}
\toprule 
\multirow{2}{*}{Parliament Elections 2018} & \multirow{2}{2cm}{\centering{}Statistic} & \multirow{2}{2cm}{\centering{}Entropy} & \multirow{2}{2cm}{\centering{}Relative Entropy} & \multirow{2}{2cm}{\centering{}Tsallis Entropy} & \multirow{2}{2.5cm}{\centering{}Relative Tsallis Entropy}\tabularnewline\addlinespace[0.2cm]
 &  &  &  &  & \tabularnewline\addlinespace[0.2cm]
\midrule
\midrule 
\multirow{2}{*}{Adjacency auto-correlation} & \centering{}Moran I & \centering{}$0.473$ & \centering{}$0.464$ & \centering{}$0.551$ & \centering{}$0.526$\tabularnewline\addlinespace[0.2cm]
\cmidrule{2-6} \cmidrule{3-6} \cmidrule{4-6} \cmidrule{5-6} \cmidrule{6-6} 
 & \centering{}p-value & \centering{}$9.46\times10^{-5}$ & \centering{}$1.22\times10^{-4}$ & \centering{}$8.56\times10^{-6}$ & \centering{}$1.90\times10^{-6}$\tabularnewline\addlinespace[0.2cm]
\midrule 
\multirow{2}{*}{Gravitational auto-correlation} & \centering{}Moran I & \centering{}$0.406$ & \centering{}$0.289$ & \centering{}$0.437$ & \centering{}$0.375$\tabularnewline\addlinespace[0.2cm]
\cmidrule{2-6} \cmidrule{3-6} \cmidrule{4-6} \cmidrule{5-6} \cmidrule{6-6} 
 & \centering{}p-value & \centering{}$6.15\times10^{-6}$ & \centering{}$6.32\times10^{-4}$ & \centering{}$1.41\times10^{-6}$ & \centering{}$2.37\times10^{-5}$\tabularnewline\addlinespace[0.2cm]
\midrule 
\multirow{2}{*}{Sectarian auto-correlation} & \centering{}Moran I & \centering{}$0.683$ & \centering{}$0.712$ & \centering{}$0.792$ & \centering{}$0.804$\tabularnewline\addlinespace[0.2cm]
\cmidrule{2-6} \cmidrule{3-6} \cmidrule{4-6} \cmidrule{5-6} \cmidrule{6-6} 
 & \centering{}p-value & \centering{}$1.36\times10^{-9}$ & \centering{}$3.06\times10^{-10}$ & \centering{}$3.79\times10^{-12}$ & \centering{}$1.84\times10^{-12}$\tabularnewline\addlinespace[0.2cm]
\bottomrule
\end{tabular}\caption{Spatial adjacency, gravitational and sectarian auto-correlation of
entropy, relative entropy, Tsallis entropy and relative Tsallis entropy
among Lebanese electoral districts in the May 2018 Parliamentary elections
with their p-values, using the Moran I index. }
\end{table}
\par\end{center}

We consider two consecutive elections held in district $j$, and we designate the number of lists competing in the two elections in that district by $i$ and $k$ respectively. We define $x_{ji}$to be the fraction of votes gained in the first election by list $i$ in district
$j$ and $y_{jk}$ to be its counterpart in the following elections.
We obviously have the condition that:

\begin{equation}
\sum_{i}x_{ji}=\sum_{k}y_{jk}=1\ \forall j
\end{equation}

To study the voter transitions between lists, we need to estimate
the Markov transition matrix $\boldsymbol{p}_{j}$ whose elements
$p_{j,ik}$ measure the probability that an individual voted for list $k$ in the last elections given that he/she has voted for list $i$ in the previous elections. This is a pure Markov inverse
problem. It is solved by maximizing the entropy of voter migration defined by:

\begin{equation}
H(\boldsymbol{p}_{j})=-\sum_{i}\sum_{k}p_{j,ik}\log(p_{j,ik})
\end{equation}

subject to several conditions that include (13), and the following
constraints:

\begin{equation}
\begin{cases}
y_{jk}=\sum_{i}x_{ji}p_{j,ik} & \forall j\forall k\\
\sum_{k}p_{j,ik}=1 & \forall j\forall i\\
p_{j,ik}\geq0 & \forall j\forall i\forall k
\end{cases}
\end{equation}

In addition to to these constraints, any information available from
exit polls or trusted estimates can be imposed on $p_{j,ik}$ as additional
constraints, and would improve the accuracy of this estimate. The
model is a maximization problem with constraints, and can be solved
by Lagrangian methods. However, closed form analytic solution are
not available, so the problem is solved numerically.

\section{Results and discussion}

\subsection{Entropic Auto-correlation}

The $2018$ Lebanese parliamentary elections were held on May $6,2018$,
and were the first parliamentary elections to be held in the country
in nine years \cite{Parties2018}. The elections were held under
a new proportional representation electoral law with one preferential
vote, which aimed to make the electoral process more inclusive and
transparent. The country is divided into $15$ electoral districts,
which contain $26$ sub-districts. The lists run on the district level
and the seats are allocated proportionally with respect to their votes
in these districts, but simultaneously reserving fixed quotas for
the sub-districts and for the sectarian constituents of the districts.
In this sense, all seats across all districts are allocated according
to a predefined sectarian distribution. The preferential votes are
used to rank the candidates of all lists that attain the minimum threshold
for winning, then the winning candidates are declared in decreasing
rank, up until the sectarian quota of each sect is fulfilled. Once
a sectarian quota is reached, all candidates from the same sect are
eliminated and the process continues until all the dedicated seats
of the winning lists are reached. The electoral threshold is set to
be the percentage of votes needed to achieve one whole seat. In this
context, it varies from one district to another, varying between $20\%$
in the smallest district that is represented by $5$ seats and $7.7\%$
in the largest district represented by $13$ seats.

The main political forces that participated in the $2018$ Lebanese
elections were the sectarian parties that controlled the power during
the last three decades: The Free Patriotic Current (FPC), The Future
Movement (FM), The Lebanese Forces (LF), Hizbullah, Amal (Hope) Movement
and the Progressive Socialist Party (PSP), in addition to minor allies
of them and several opposition parties and groups which couldn't break
their hegemony. The coalition of March $8$, including the FPC, Hizbullah
and Amal held the majority with some of their minor allies.
\begin{center}
\begin{table}
\centering{}%
\begin{tabular}{c>{\centering}b{2cm}>{\centering}m{2cm}>{\centering}m{2cm}>{\centering}m{2cm}>{\centering}m{2.5cm}}
\toprule 
\multirow{2}{*}{Parliament Elections 2022} & \multirow{2}{2cm}{\centering{}Statistic} & \multirow{2}{2cm}{\centering{}Entropy} & \multirow{2}{2cm}{\centering{}Relative Entropy} & \multirow{2}{2cm}{\centering{}Tsallis Entropy} & \multirow{2}{2.5cm}{\centering{}Relative Tsallis Entropy}\tabularnewline\addlinespace[0.2cm]
 &  &  &  &  & \tabularnewline\addlinespace[0.2cm]
\midrule
\midrule 
\multirow{2}{*}{Adjacency auto-correlation} & \centering{}Moran I & \centering{}$0.494$ & \centering{}$0.402$ & \centering{}$0.529$ & \centering{}$0.503$\tabularnewline\addlinespace[0.2cm]
\cmidrule{2-6} \cmidrule{3-6} \cmidrule{4-6} \cmidrule{5-6} \cmidrule{6-6} 
 & \centering{}p-value & \centering{}$5.21\times10^{-5}$ & \centering{}$6.55\times10^{-4}$ & \centering{}$1.76\times10^{-5}$ & \centering{}$3.96\times10^{-4}$\tabularnewline\addlinespace[0.2cm]
\midrule 
\multirow{2}{*}{Gravitational auto-correlation} & \centering{}Moran I & \centering{}$0.491$ & \centering{}$0.416$ & \centering{}$0.490$ & \centering{}$0.453$\tabularnewline\addlinespace[0.2cm]
\cmidrule{2-6} \cmidrule{3-6} \cmidrule{4-6} \cmidrule{5-6} \cmidrule{6-6} 
 & \centering{}p-value & \centering{}$8.10\times10^{-8}$ & \centering{}$3.36\times10^{-6}$ & \centering{}$8.45\times10^{-8}$ & \centering{}$5.57\times10^{-7}$\tabularnewline\addlinespace[0.2cm]
\midrule 
\multirow{2}{*}{Sectarian auto-correlation} & \centering{}Moran I & \centering{}$0.760$ & \centering{}$0.665$ & \centering{}$0.846$ & \centering{}$0.804$\tabularnewline\addlinespace[0.2cm]
\cmidrule{2-6} \cmidrule{3-6} \cmidrule{4-6} \cmidrule{5-6} \cmidrule{6-6} 
 & \centering{}p-value & \centering{}$2.32\times10^{-11}$ & \centering{}$3.22\times10^{-9}$ & \centering{}$1.53\times10^{-13}$ & \centering{}$1.91\times10^{-12}$\tabularnewline\addlinespace[0.2cm]
\bottomrule
\end{tabular}\caption{Spatial adjacency, gravitational and sectarian auto-correlation of
entropy, relative entropy, Tsallis entropy and relative Tsallis entropy
among Lebanese electoral districts in the May 2022 Parliamentary elections
with their p-values, using the Moran I index.}
\end{table}
\par\end{center}

The $2022$ elections were held on May $15,2022$ under the same proportional
confessional electoral law. The main parties mentioned above were
challenged this time by lists that were based on coalition between
new and emerging opposition force and existing secular forces. The
elections resulted in a hung parliament were March $8$ coalition
was still the most numerous group but without a majority \cite{Parties2022,Parties2022-1}.
About $19$ seats were won by candidates of the secular opposition
lists.

In both elections, most parties ran in coalition lists with other
parties and independent candidates, and the coalitions varied from
one district to another, ranging from political to pragmatic coalitions
to guarantee winning seats under the existing proportional law.

Figure (1) summarizes the values of Shannon's entropy, relative Shannon's
entropy, Tsallis entropy and relative Tsallis entropy of entropic
index $2$ on the level of the electoral district in the elections
of $2018$ and $2022$. They are calculated using equations (1), (3),
(4) and (6). We can notice that, for most districts, there was an
increase in entropy between the two elections. 

In order to better understand possible correlation between different
districts, we applied the Moran I index tool as defined in (9) to
measure the auto-correlation between local district voter entropies
with geographical district adjacency, voter gravitational adjacency
and sectarian adjacency using the results of the $2018$ and the $2022$
election results that publicly available in \cite{LebElec4,LebElec5,LebElec6}.
We found that there is a positive auto-correlation among all those
values with confidence level greater than $99.9\%$ for all measured
Moran indices. This means that we can reject the null hypothesis with
a very high confidence level in all these cases, and quantify the
positive correlations between the entropies and the analyzed geographic,
demographic and political features.

Table (1) shows that in the $2018$ Lebanese parliamentary elections,
there is strong entropy-adjacency auto-correlation, where the Tsallis
entropy is the most auto-correlated with adjacency. This means that
geographically neighboring districts have correlated patterns of voter
distribution uncertainty. Gravitational auto-correlation is positive,
but is weaker than adjacency for all entropic measures. This indicates
that the effect of the ratio of voter population to the inverse of
the squared distance between districts is weaker than that of geographic
adjacency. Finally, the strongest auto-correlation occurs for sectarian
adjacency, meaning that districts with similar sectarian majorities
will have the strongest correlation between their voter distributions
uncertainty, reflecting the dominant weight of the sectarian identity
in Lebanese politics.

In the $2022$ Lebanese parliamentary elections, there also was a
strong relationship between the distribution of voter uncertainty
and geographic proximity, with neighboring districts showing correlated
patterns in this regard, as shown in Table (2). The impact of the
ratio of voter population to the inverse of the squared distance between
districts was similar to that of geographic proximity, hence we could
see a stronger pattern of geographic auto-correlation in this elections
that in the previous one. Sectarian similarity again had the strongest
influence on the correlation between voter uncertainty in different
districts, with districts with similar sectarian majorities showing
the strongest correlations in this regard. This reflects the continued
significant role that sectarian identity plays in Lebanese politics.

Our findings about geographic adjacency and proximity are in line with the outcomes of other studies that analyzed spatial correlation of turnout rates and found out that the fraction of winning votes decays logarithmically with the distance between towns \cite{Borghesi}, using a different theoretical framework.

Sectarianism refers to the division of society or politics along religious
or ethnic lines. It can have a significant impact on voting patterns
because people may be more likely to vote for candidates or parties
that align with their religious or ethnic identity. This can lead
to a situation where certain groups are more likely to support certain
political parties or candidates, which can create divisions within
a society. Sectarianism can also lead to political polarization and
can make it more difficult for politicians to build coalitions or
reach compromise. It can also create an environment where politicians
feel pressure to take positions or adopt policies that appeal to their
particular sectarian group, rather than considering the interests
of the entire society.
\begin{center}
\begin{table}
\begin{centering}
\begin{tabular}{>{\raggedright}m{3cm}>{\raggedright}m{1.5cm}>{\raggedright}m{1.55cm}>{\raggedright}m{1.5cm}>{\raggedright}m{1.7cm}>{\raggedright}m{1.55cm}>{\raggedright}m{1.2cm}>{\raggedright}m{1.3cm}}
\toprule 
\centering{}\textbf{Origin 2018} & \multicolumn{7}{c}{\textbf{Beirut II Voter Destination 2022}}\tabularnewline\addlinespace
\midrule
\midrule 
\multirow{2}{3cm}{\centering{}\textbf{List Name}} & \multirow{2}{1.5cm}{\centering{}Beirut Unity} & \multirow{2}{1.55cm}{\centering{}Beirut for Change} & \multirow{2}{1.5cm}{\centering{}This is Beirut} & \multirow{2}{1.7cm}{\centering{}Bei. needs a Heart} & \multirow{2}{1.55cm}{\centering{}Beirut Confronts} & \multirow{2}{1.2cm}{\centering{}Others} & \multirow{2}{1.3cm}{\centering{}Non Voters}\tabularnewline\addlinespace
 &  &  &  &  &  &  & \tabularnewline\addlinespace
\midrule 
\centering{}Future for Beirut & \centering{}$0.015\%$ & \centering{}$21.2\%$ & \centering{}$16.0\%$ & \centering{}$5.13\%$ & \centering{}$16.6\%$ & \centering{}$17.6\%$ & \centering{}$23.5\%$\tabularnewline\addlinespace
\midrule 
\centering{}Beirut Unity & \centering{}$74.4\%$ & \centering{}$5.28\%$ & \centering{}$4.13\%$ & \centering{}$1.65\%$ & \centering{}$4.26\%$ & \centering{}$4.48\%$ & \centering{}$5.81\%$\tabularnewline\addlinespace
\midrule 
\centering{}Lebanon's Worth it & \centering{}$0.434\%$ & \centering{}$2.76\%$ & \centering{}$2.54\%$ & \centering{}$86.2\%$ & \centering{}$2.57\%$ & \centering{}$2.61\%$ & \centering{}$2.85\%$\tabularnewline\addlinespace
\midrule 
\centering{}Beirut is Homeland & \centering{}$3.99\%$ & \centering{}$9.19\%$ & \centering{}$51.9\%$ & \centering{}$7.72\%$ & \centering{}$8.90\%$ & \centering{}$8.97\%$ & \centering{}$9.32\%$\tabularnewline\addlinespace
\midrule 
\centering{}All for Beirut & \centering{}$1.24\%$ & \centering{}$86.5\%$ & \centering{}$2.48\%$ & \centering{}$2.19\%$ & \centering{}$2.49\%$ & \centering{}$2.50\%$ & \centering{}$2.59\%$\tabularnewline\addlinespace
\midrule 
\centering{}Others & \centering{}$10.8\%$ & \centering{}$15.1\%$ & \centering{}$14.9\%$ & \centering{}$14.1\%$ & \centering{}$14.9\%$ & \centering{}$15.0\%$ & \centering{}$15.2\%$\tabularnewline\addlinespace
\midrule 
\centering{}Non-Voters & \centering{}$0.00\%$ & \centering{}$4.17\%$ & \centering{}$1.32\%$ & \centering{}$0.018\%$ & \centering{}$1.53\%$ & \centering{}$1.94\%$ & \centering{}$91.0\%$\tabularnewline\addlinespace
\bottomrule
\end{tabular}
\par\end{centering}
\caption{GME estimate of the voter transitions between lists in Beirut II district
between May 2018 and May 2022 parliamentary elections.}
\end{table}
\par\end{center}

\subsection{Voter Transitions}

Changes in the voting patterns of citizens during elections can be affected by multiple factors, and the significance of each factor may vary based on the election and the electorate's preferences. In the case of the Lebanese parliamentary elections, various factors, such as significant economic, social, security, and political developments that happened in Lebanon from $2018$ to $2022$, might have contributed to the changes in voting behavior of the people.
Lebanon has been grappling with a severe economic crisis since $2019$, which has caused an economic collapse. The country also witnessed a large-scale uprising between $2019$ and $2020$, with significant protests expressing widespread dissatisfaction \cite{Parties2022}. Former Prime Minister Saad Hariri resigned, left the country, and subsequently withdrew from politics, including not running in the $2022$ elections \cite{Hariri}. The $2020$ explosion at the Beirut port had devastating consequences for the city and its surrounding areas, directly impacting the lives of hundreds of thousands of residents \cite{Port}. Security and economic factors are important variables that can affect election results in various ways. In particular, they can influence voter behavior, such as turnout, and campaign messaging. In this context, the aforementioned developments could have influenced voter transitions, possibly into opposition lists, in several districts.  In addition, the results obtained in Figure (1) reveal a small yet observable increase in entropy in $2022$, indicating a more uniform spread of votes away from dominant lists. These qualitative and quantitative factors combined could provide a logical validation for the actual voter transitions between $2018$ and $2022$ elections.

We apply the GME estimation method to determine the voter transitions
between lists in the Lebanese parliamentary elections of $2018$ and
$2022$. We show in this paper the corresponding voter transitions
in $3$ electoral districts with different sectarian and political
compositions. In particular, we choose to show the voter transitionss
in the districts of Beirut II, South III and Mount Lebanon IV.

\subsubsection*{Beirut II}

The Lebanese capital Beirut is divided into two electoral districts
in the current election law adopted since $2018$. The second district
(Beirut II) has $11$ seats, distributed according to a sectarian
quota that assigns $6$ seats for Muslim Sunnis, $2$ seats for Muslim
Shia, $1$ seat for Druze, $1$ seat for Christian Orthodox and $1$
seat for Christian evangelicals. The main political parties in this
district are the Future Movement, Hizbullah, Amal movement, the Progressive
Socialist Party, Free Patriotic Current, and several other smaller
sectarian sectarian parties. In addition, a coalition of opposition
secular forces established itself as an alternative force, especially
in the $2022$ elections. In $2018,$ the main electoral lists were:
"Future for Beirut'', a coalition between the Future movement,
Progressive Socialist Party and other minor groups; "Beirut Unity'',
a coalition between Hizbullah, Amal movement, Free patriotic current,
the Islamic Project Association and other smaller groups; and several
other lists. In $2022$, the Future movement boycotted the elections,
and several lists, mainly "This is Beirut'' and "Beirut Confronts''
were formed by its formal members and allies. The "Beirut Unity''
continued, with the Islamic Project Association forming its own list.
"Lebanon's Worth it'' list of the former elections re-branded itself
as "Beirut needs Heart''. The main challenge came from the new
secular opposition list, "Beirut for Change''.

The GME estimator shows that the $2018$ voters of "Future for Beirut''
migrated in $2022$ elections into non-voters ($23.5\%)$, "Beirut
for Change'' ($21.2\%$), Others ($17.6\%$), "Beirut Confronts''
($16.6\%$) and "This is Beirut'' ($16\%$). On the other hand,
$74.4\%$ of the voters of "Beirut Unity'' were retained by the
list, while the others votes migrated to the other list. "Beirut
needs a heart'' retained $86.2\%$ of its former voters of "Lebanon's
Worth it'' list. "Beirut for Change'' also inherited about $86.5\%$
of the votes of "All for Beirut'', a secular list of the $2018$
elections. $91\%$ of non-voters from the previous elections remained
as non-voters, while around $4.2\%$ of them voted for "Beirut for
Change'' in the $2022$ elections. The full transition estimates
are shown in Table 3.

\subsubsection*{Mount Lebanon IV}

Mount Lebanon IV is the district with highest number of representatives
in the Lebanese parliament, with $13$ seats. It is represented according
to the sectarian quota: $5$ seats for Christian Maronites, $1$ seat
for Christian Catholics, $1$ seat for Christian Orthodox, $4$ seats
for Druze and $2$ seats for Muslim Sunni. The main political parties
in the district are the Progressive Socialist Party, the Lebanese
Forces, the Free Patriotic Movement, the Lebanese Democratic Party,
Future Movement, the Arab Unity Party and other smaller parties and
groups, in addition to a broad spectrum of secular opposition parties
and groups. In the $2018$ elections, the main lists were the "Reconciliation''
list, formed by the coalition of the Progressive Socialist Party with
the Future Movement and the Lebanese Forces; "Mountain Guarantee''
list formed by the Lebanese Democratic Party, the Free Patriotic Current
and their allies; "National Unity'' formed by the Arab Unity Party;
and "All for my Country'' which was a coalition of secular and
left forces. In $2022$, the former components of the "Mountain
Guarantee'' and "National Unity'' list merged in one list called
"The Mountain'' list, while the constituents of the "Reconciliation''
list formed the "Will and Participation'' list, retaining all of
its components except the Future Movement. All secular, left and civil
opposition forces collaborated in a broad alliance list that they
called "Unified for Change''.

The GME estimates shown in Table 4 reveal that $81.2\%$ of the voters
of the "Reconciliation'' list of $2018$ voted for the "Will
and Participation'' in $2022$, while about $17\%$ of its voters
migrated their votes to "Unified for Change''. About $80\%$ of
the votes of "Mountain Guarantee'' and "National Unity'' were
retained by the merger list "The Mountain'' while $18.2\%$ and
$11.4\%$ of their votes respectively migrated to the "Unified for
Change'' list. About "$93.6\%$ of the voters of "All for my
Country'' were retained by the "Unified for Change'' list. The
vast majority ($96.4\%$) of non-voters in $2018$ did not vote in
$2022$ as well, while about $2.5\%$ of these former non-voters voted
for "Unified for Change''.
\begin{center}
\begin{table}
\begin{centering}
\begin{tabular}{>{\raggedright}m{3.2cm}>{\raggedright}m{2.1cm}>{\raggedright}m{2cm}>{\raggedright}m{1.8cm}>{\raggedright}m{1.4cm}>{\raggedright}m{1.4cm}}
\toprule 
\centering{}\textbf{Origin 2018} & \multicolumn{5}{c}{\textbf{Mount Lebanon IV Voter Destination 2022}}\tabularnewline\addlinespace
\midrule
\midrule 
\multirow{2}{3.2cm}{\centering{}\textbf{List Name}} & \multirow{2}{2.1cm}{\centering{}Will \& Participation} & \multirow{2}{2cm}{\centering{}Unified for Change} & \multirow{2}{1.8cm}{\centering{}The Mountain} & \multirow{2}{1.4cm}{\centering{}Others} & \multirow{2}{1.4cm}{\centering{}Non Voters}\tabularnewline\addlinespace
 &  &  &  &  & \tabularnewline\addlinespace
\midrule 
\centering{}Reconciliation & \centering{}$81.2\%$ & \centering{}$16.9\%$ & \centering{}$1.48\%$ & \centering{}$0.24\%$ & \centering{}$0.11\%$\tabularnewline\addlinespace
\midrule 
\centering{}Mountain Guarantee & \centering{}$1.77\%$ & \centering{}$18.2\%$ & \centering{}$79.6\%$ & \centering{}$0.19\%$ & \centering{}$0.23\%$\tabularnewline\addlinespace
\midrule 
\centering{}National Unity & \centering{}$5.31\%$ & \centering{}$11.4\%$ & \centering{}$79.8\%$ & \centering{}$2.17\%$ & \centering{}$1.33\%$\tabularnewline\addlinespace
\midrule 
\centering{}All for my Country & \centering{}$2.28\%$ & \centering{}$93.6\%$ & \centering{}$3.01\%$ & \centering{}$1.13\%$ & \centering{}$0.76\%$\tabularnewline\addlinespace
\midrule 
\centering{}Others & \centering{}$20.7\%$ & \centering{}$31.0\%$ & \centering{}$26.4\%$ & \centering{}$12.5\%$ & \centering{}$9.33\%$\tabularnewline\addlinespace
\midrule 
\centering{}Non-Voters & \centering{}$0.08\%$ & \centering{}$2.51\%$ & \centering{}$0.108\%$ & \centering{}$0.01\%$ & \centering{}$96.4\%$\tabularnewline\addlinespace
\bottomrule
\end{tabular}
\par\end{centering}
\caption{GME estimate of the voter transitions between lists in Mount Lebanon
IV district between May 2018 and May 2022 parliamentary elections.}
\end{table}
\par\end{center}

\subsubsection*{South III}

South III electoral district is represented by $11$ members of parliament
that are assigned according to the following sectarian quota: $8$
seats for Muslim Shia, $1$seat for Muslim Sunni, $1$ seat for Druze
and $1$ seat for Christian Orthodox. The main political parties are
Hizbullah, Amal Movement, the Lebanese Communist Party, Future movement,
Lebanese Forces and Free Patriotic Current, in addition to other minor
parties and groups. In the $2018$ elections, the main lists were
the "Hope and Loyalty'', a coalition between Hizbullah, Amal movement
and other minor forces; "The South Deserves'', a coalition between
Future movement and the Free Patriotic Current; "One Vote for Change'',
formed by the Lebanese Communist Party; and other minor lists. In
the $2022$ elections, "Hope and Loyalty'' continued as an alliance
list of all of its former constituents, while all secular and progressive
forces and groups formed the "Together for Change'' list in alliance
with the Lebanese Communist Party. The Future movement boycotted the
elections, while the Free Patriotic Current and the Lebanese Forces
did not take part in any list. 

Based on the GME estimator method, we show in Table 5 the voter transitions
among these lists between the $2018$ and $2022$ elections. We find
out that the "Hope and Loyalty'' list retained $96.6\%$ of its
voters, while only about $1.7\%$ of its former voters supported "Together
for Change'' in $2022$. On the other hand, "The South Deserves''
saw most of its voters supporting "Together for Change'' in $2022$,
while about $19.7\%$ of its voters transitioned into non-voting,
in line with the call of the Future Movement for election boycott.
Most of the voters of "One Vote for Change'' voted for "Together
for Change'' in $2022$, with a transition estimated at $95.3\%$,
while more than $3\%$ of its former voters became non-voters in $2022$.
We also estimate that the majority of the voters of other lists in
$2018$ either voted for "Together for Change'' or did not vote.
More than $96\%$ of non-voters in $2018$ did not vote in $2022$
while about $2.1\%$ of them voted for "Hope and Loyalty'' and
$1.4\%$ for "Together for Change''.
\begin{center}
\begin{table}
\begin{centering}
\begin{tabular}{>{\raggedright}m{3.4cm}>{\raggedright}m{2.1cm}>{\raggedright}m{2cm}>{\raggedright}m{1.8cm}>{\raggedright}m{1.4cm}}
\toprule 
\centering{}\textbf{Origin 2018} & \multicolumn{4}{c}{\textbf{South III Voter Destination 2022}}\tabularnewline\addlinespace
\midrule
\midrule 
\multirow{2}{3.4cm}{\centering{}\textbf{List Name}} & \multirow{2}{2.1cm}{\centering{}Hope \& Loyalty} & \multirow{2}{2cm}{\centering{}Together for Change} & \multirow{2}{1.8cm}{\centering{}Voice of South} & \multirow{2}{1.4cm}{\centering{}Non Voters}\tabularnewline\addlinespace
 &  &  &  & \tabularnewline\addlinespace
\midrule 
\centering{}Hope \& Loyalty & \centering{}$96.6\%$ & \centering{}$1.73\%$ & \centering{}$0.049\%$ & \centering{}$0.51\%$\tabularnewline\addlinespace
\midrule 
\centering{}South Deserves & \centering{}$8.31\%$ & \centering{}$71.7\%$ & \centering{}$0.050\%$ & \centering{}$19.7\%$\tabularnewline\addlinespace
\midrule 
\centering{}One Vote for Change & \centering{}$0.92\%$ & \centering{}$95.3\%$ & \centering{}$0.23\%$ & \centering{}$3.35\%$\tabularnewline\addlinespace
\midrule 
\centering{}Fed up with Rhetoric & \centering{}$5.11\%$ & \centering{}$62.6\%$ & \centering{}$0.25\%$ & \centering{}$32.2\%$\tabularnewline\addlinespace
\midrule 
\centering{}Others & \centering{}$6.96\%$ & \centering{}$84.9\%$ & \centering{}$0.24\%$ & \centering{}$8.1\%$\tabularnewline\addlinespace
\midrule 
\centering{}Non-Voters & \centering{}$2.11\%$ & \centering{}$1.41\%$ & \centering{}$0.049\%$ & \centering{}$96.2\%$\tabularnewline\addlinespace
\bottomrule
\end{tabular}
\par\end{centering}
\caption{GME estimate of the voter transitions between lists in South III district
between May 2018 and May 2022 parliamentary elections.}
\end{table}
\par\end{center}

\subsection{Discussion and Limitations}

In our analysis of voter transitions, we considered the percentage
of votes attained by each list, without reflecting that on the total
number of votes. One of the main issues in such analysis is the number
of voters that varies from one election to another due to inflows
of new eligible voters mainly due to young people reaching the voting
age and naturalized citizen, and outflows of dead or emigrating voters.
The change in the total number of voters puts some challenges on the
interpretation of transition probabilities, that mainly represent
the probabilities of changing an electoral preference from one election
to another. However, the inflows and outflows are not taken into account
in such an analysis. In this paper, we have assumed the total voting
population to be fixed, and we ignored the small relative changes
in this population over a course of four years. 

The second point of interest is that the percentage of non-voters
is relatively very high in Lebanon due to the fact that high proportion
of the enlisted voters on the official records may have already left
the country for good. Recent estimates put the number of Lebanese
emigrants at around one third of the population, but they would remain
enlisted on the records \cite{Immigration, Immigration2}, leading
to participation rates varying historically around $40-50\%$ of the
registered voters on the national level \cite{LebElec4, LebElec5, LebElec6}.
This complicates the question of accurately determining the actual
number of registered voters who might eventually want to vote in the
elections. For practical purposes, we have classified all the non-voting
population as non-voters regardless of their residence/immigration
statuses, in addition to all blank and invalid votes, taken together
as a single block.

\section{Conclusions}

In this paper, we defined different measures of entropy, namely the
Shannon, relative Shannon, Tsallis and relative Tsallis entropies
of entropic index two. We introduced the Moran I index and used it
to measure the spatial auto-correlation of these entropies with respect
to geographic adjacency, gravitational adjacency and sectarian adjacency.
The various adjacency features were quantified through the adjacency
matrix whose elements are defined as: non-zero when two districts
are geographically adjacent, the product of the number of voters in
each two districts over the square of the distance between their centers,
and non-zero when two districts have same sectarian majority, respectively.
This is the main novel model of our paper, and it allowed us to study
the corresponding correlations between entropies of election outcomes
on the district level. We found out that, with a very high degree
of confidence, there are relatively strong geographic and gravitational
adjacency correlations, but most importantly and most strongly, there
is a very apparent auto-correlation in the sectarian adjacency feature,
in both parliamentary elections in Lebanon in $2018$ and $2022$.
The result presents an important assertion on the high impact of the
sectarian effect in Lebanese politics and elections. 

We also introduced the method of the maximized general entropy estimation
that allows us to determine the Markov transition matrix of list voters
between consecutive elections. We employed this method to determine
the voting transitions in three different key electoral districts
between the $2018$ and $2022$ Lebanese parliamentary elections.
We also pointed out some limitations and related assumptions.

Without loss of generality, we applied the model on data from Lebanese
elections, but they are universally valid. Future work can extend
the entropic spatial auto-correlation model to cover additional features
of interest, and applications on other elections.

\section*{Acknowledgments}

The author thanks Werner Antweiler for his useful comments about the
literature on the maximum entropy estimation of transition probabilities
and related software packages for its calculation. 

\section*{Appendix}

The $z_{I}$-score is defined to be:

\[
z_{I}=\frac{I-E[I]}{\sqrt{V[I]}}
\]

where $E[I]$ is the expected value and $V[I]$ is the variance. The
expected value of Moran's index is explicitly given by 
\[
E[I]=\frac{-1}{N-1}
\]

while the variance is 

\[
V[I]=E[I^{2}]-E[I]^{2}
\]

with 
\[
E[I^{2}]=\frac{A-B}{C}
\]

$A$, $B$ and $C$ are given by:

\[
A=N\left[2\left(N^{2}-3N+3\right)\Sigma_{ij}W_{ij}^{2}-2N\Sigma_{i}\left(\Sigma_{j}W_{ij}\right)^{2}+3\left(\Sigma_{ij}W_{ij}\right)^{2}\right]
\]

\[
B=\frac{2\Sigma_{i}(X_{i}-\bar{X})^{4}}{\left(\Sigma_{i}(X_{i}-\bar{X})^{2}\right)^{2}}\left[\left(N^{2}-N\right)\Sigma_{ij}W_{ij}^{2}-2N\Sigma_{i}\left(\Sigma_{j}W_{ij}\right)^{2}+3\left(\Sigma_{ij}W_{ij}\right)^{2}\right]
\]

\[
C=\left(N-1\right)\left(N-2\right)\left(N-3\right)\left(\Sigma_{ij}W_{ij}\right)^{2}
\]

\bibliographystyle{elsarticle-num}

\end{document}